\documentclass[twocolumn,showpacs,preprintnumbers,amsmath,amssymb]{revtex4}
\usepackage{graphicx}
\usepackage{dcolumn}
\usepackage{bm}

\begin{document}
\title{Associated production of a neutral top-Higgs with a heavy-quark pair in the $\gamma\gamma$ collisions at ILC}
\author{Jinshu Huang}
\email{jshuang@vip.sina.com} \affiliation{College of Physics $\&$
Information Engineering, Henan Normal University, Xinxiang 453007,
People's Republic of China; \\ College of Physics $\&$ Electric
Engineering, Nanyang Normal University, Nanyang 473061, People's
Republic of China}
\author{Gongru Lu}
\email{lugongru@sina.com}
\author{Wei Xu}
\affiliation{College of Physics $\&$ Information Engineering, Henan
Normal University, Xinxiang 453007, People's Republic of China}
\author{Shuaiwei Wang}
\affiliation{College of Physics $\&$ Electric Engineering, Nanyang
Normal University, Nanyang 473061, People's Republic of China}

\date{\today}

\begin{abstract}
We have studied the associated production processes of a neutral
top-Higgs in the topcolor assisted technicolor model with a pair of
heavy quarks in $\gamma\gamma$ collisions at the International
Linear Collider (ILC). We find that the cross section for
$t\bar{t}h_t$ in $\gamma\gamma$ collisions is at the level of a few
${\rm fb}$ with the c.m. energy $\sqrt{s}=1000$ GeV, which is
consistent with the results of the cross section of $t\bar{t}H$ in
the standard model and the cross section of $t\bar{t}h$ in the
minimal supersymmetric standard model and the littlest Higgs models.
It should be clear that hundreds of to thousands of $h_t$ per year
can be produced at the ILC. This process of $\gamma\gamma
\rightarrow t\bar{t}h_t$ is really interesting in testing the
standard model and searching the signs of technicolor.
\end{abstract}


\maketitle

\section{\label{sec:level1}Introduction}

\ \ \ The electroweak symmetry breaking (EWSB) mechanism remains an
open question in spite of the success of the standard model (SM)
compared with the precision measurement data. With the advent of the
new collider technique, high energy and high intensity photon beams
can be obtained by using Compton laser photons scattering off the
colliding electron and positron beams \cite{Ginzburg1981}. The
collisions of high energy photons produced at the linear collider
provide a comprehensive laboratory for testing the SM and probing
new physics beyond the SM \cite{Brodsky1995}.

As we know, the initial technicolor (TC) \cite{Weinberg1976}, as a
theory of dynamical EWSB, is one of the important candidates for new
physics beyond the SM, especially the topcolor assisted technicolor
(TC2) model proposed by C. T. Hill \cite{Hill1995}. This combines
technicolor with topcolor, with the former mainly responsible for
EWSB and the latter for generating a major part of the top quark
mass. Since this model could provide rational answers to some of the
questions, it is of significant interest. This model predicts three
top-pions ($\pi^0_t, \pi^{\pm}_t$) and one top-Higgs ($h_t$) with
large Yukawa couplings to the third generation quarks, so these new
particles can be regarded as a typical feature of the model. Many
signals of the model have already been studied in the work
environment of linear colliders and hadron hadron colliders
\cite{Cao2003,Yue2002, Wang2003-1}, but much of the attention was
focused on the neutral and charged top pions and new gauge bosons.
Here we wish to discuss the prospects of neutral top-Higgs.

In the SM, the Higgs boson associated production with a pair of top
quarks in the high energy photon collisions has been calculated
\cite{Cheung1993}, and Reference \cite{Gunion1993} presents a study
of the process $\gamma\gamma \rightarrow t\overline{t}\phi$
$(\phi=h^{0},H^{0},A^{0})$ in the minimal supersymmetric standard
model (MSSM). The authors of Ref. \cite{Pan2007} investigated the
associated $t\bar{t}h^0$ production process $e^+e^- \rightarrow
\gamma\gamma \rightarrow t\bar{t}h^0$ at the future $e^+e^-$ linear
colliders up to QCD next-to-leading order in the frameworks of the
littlest Higgs (LH) model and its extension with T-parity (LHT). In
Ref. \cite{Wang2003-2}, the authors have calculated the associated
production of neutral top-pion with a heavy-quark pair in
$\gamma\gamma$ collisions. In this paper, we will study the
associated production of a neutral top-Higgs with a heavy-quark pair
in $\gamma \gamma$ collisions at the International Linear Collider
(ILC).

\begin{figure*}
\begin{center}

\vspace{-1.5cm}

\ \hspace{-3.3cm} \ \includegraphics{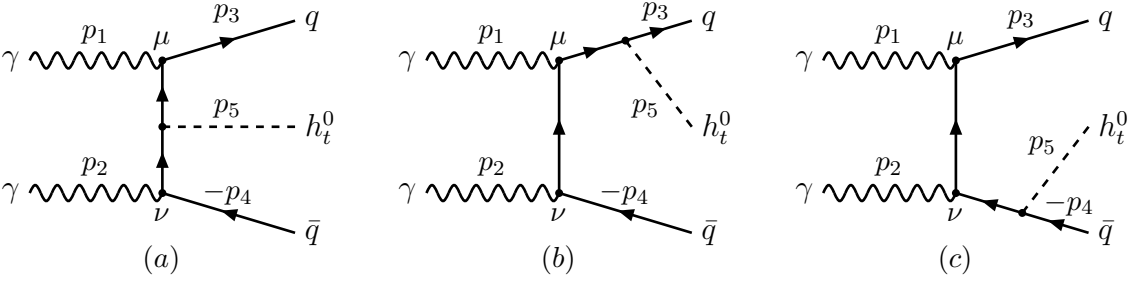}

\vspace{-23.5cm}

\caption{\label{fig:eps1} Feynman diagrams for $f\bar{f}h_t$
associated production in $\gamma\gamma$ collisions. It is not
plotted for those diagrams showing the interchange of the two
incoming photons.}

\end{center}
\end{figure*}

\section{\label{sec:level2} The cross section of $f\bar{f}h_t$ in the high energy $\gamma\gamma$ collisions}

\ \ \ As a rough estimate, we only consider the process
$\gamma(p_{1}) \gamma(p_{2}) \rightarrow f(p_{3} )\bar{f}(p_{4})
h_{t}(p_{5})$ $(f=t,b)$ at the tree level. The Feynman diagrams are
shown in Fig. \ref{fig:eps1}, in which those diagrams with the
interchange of the two incoming photons are not shown.

The amplitudes for this process are given by
\begin{eqnarray}
M^{(a)}_f &=& \frac{e^2 Q^2_f m^*_f}{\sqrt{2}f_{\pi_t}}
G(p_{3}-p_{1},m_{f})G(p_{2}-p_{4},m_{f}) \times \nonumber \\  &&
\bar{u}_{f}(p_{3})\not \varepsilon (p_{1}) (\not p_3- \not
p_1+m_{f})(\not p_{2}-\not p_{4} \nonumber \\  && +m_{f}) \not
\varepsilon (p_{2}) v_{f}(p_{4}),
\end{eqnarray}
\begin{eqnarray}
 M^{(b)}_f &=& \frac{e^2 Q^2_f m^*_f}{\sqrt{2}f_{\pi_t}} G(p_{3}+p_{5}, m_{f}) G(p_{2}-p_{4}, m_{f}) \times \nonumber \\
 & &
\bar{u}_{f}(p_{3}) (\not p_{3}+ \not p_5 +m_f)\not
\varepsilon(p_1)(\not p_{2}-\not p_{4} \nonumber \\  && +m_{f}) \not
\varepsilon(p_{2}) v_{f}(p_{4}),
\end{eqnarray}
\begin{eqnarray}
 M^{(c)}_f &=& \frac{e^2 Q^2_f m^*_f}{\sqrt{2}f_{\pi_t}} G(p_{3}-p_{1},m_{f})G(-p_{4}-p_{5},m_{f})\times \nonumber \\
 & &
\bar{u}_{f}(p_{3}) \not \varepsilon(p_{1})(\not p_{3}-\not
p_{1}+m_{f}) \not \varepsilon(p_{2})(-\not p_{4} \nonumber \\
&& -\not p_{5}+m_{f}) v_{f}(p_{4}).
\end{eqnarray}

The amplitudes for those diagrams with the interchange of the two
incoming photons can be directly obtained by interchanging $p_1,p_2$
in the above amplitudes. Here the subindex $f=t,b$, $m^*_t$ and
$m^*_b$ denote the masses of the top quark and bottom quark
generated by the topcolor interaction, $m^*_t=(1-\varepsilon)m_t\
(\varepsilon \approx 0.03-0.1)$, $m^*_b=k 6.6\ {\rm GeV}\ (k \approx
0.1-1)$ \cite{Hill1995}, and the  function $G$ denotes
\begin{equation}
G(p,m)=\frac{1}{p^{2}-m^{2}}.
\end{equation}

With the above amplitudes, we can directly obtain the cross section
$\hat{\sigma}(\hat{s})$ for the subprocess $\gamma \gamma
\rightarrow f \bar{f} h_{t}$, and the total cross section at the
$e^+e^-$ linear collider can be obtained by folding the elementary
cross section $\sigma(\hat{s})$ for the subprocess $\gamma\gamma
\rightarrow f\bar{f}h_t$ with the photon luminosity at the $e^+e^-$
colliders given in Ref. \cite{Gjikia1992}, i.e.,
\begin{equation}
\sigma (s)=\int^{x_{\rm max}}_{x_{\rm min}}{\rm d} x_{1}
\int^{x_{\rm max}}_{x_{\rm min} x_{\rm max}/x_1} {\rm d} x_{2}
F_{\gamma/e}(x_{1})F_{\gamma/e}(x_{2})\hat{\sigma}(\hat{s}),
\end{equation}
where $\sqrt{s}$ and $\sqrt{\hat{s}}$ are the $e^+e^-$ and
$\gamma\gamma$ center-of-mass (c.m.) energies, respectively.

For unpolarized initial electron and laser beams, the energy
spectrum of the backscattered photon is given by
 \cite{Cheung1993,Gjikia1992}
\begin{equation}
 F_{\gamma/e}(x)=\frac{1}{D(\xi)}[1-x+\frac{1}{1-x}
 -\frac{4x}{\xi(1-x)}+\frac{4x^2}{\xi^2
 (1-x^2)}],
\end{equation}
with
\begin{equation}
 D(\xi)=(1-\frac{4}{\xi}-\frac{8}{\xi^2}){\rm ln}(1+\xi) +\frac{1}{2}
 +\frac{8}{\xi}-\frac{1}{2(1+\xi)^2},
\end{equation}
where $\xi=4E_e E_0/m_e^2$, in which $m_e$ and $E_e$ denote,
respectively, the incident electron mass and energy, $E_0$ denotes
the initial laser photon energy, and $x_i=E/E_e$ is the fraction
which represents the ratio between the scattered photon and the
initial electron energy for the backscattered photons moving along
the initial electron direction. $F_{\gamma/e}(x)$ vanishes for
$x>x_{\rm max}=E_{\rm max}/E_e=\xi/(1+\xi)$. In order to avoid the
creation of $e^+e^-$ pairs by the interaction of the incident and
backscattered photons, we require $E_0 x_{\rm max} \leq m_e^2/E_e$,
which implies $\xi \leq 2+2\sqrt{2} \approx 4.8$
\cite{Cheung1993,Zhou1998}. For the choice $\xi=4.8$, it can obtain
\begin{equation}
x_{\rm max} \approx 0.83, \ \ D(\xi) \approx 1.8.
\end{equation}

The minimum value for $x$ is then determined by the production
threshold
\begin{eqnarray}
x_{\rm min}=\frac{\hat{s}_{\rm min}}{x_{\rm max}s}, \ \ \  \
\hat{s}_{\rm min}=(2m_f+m_{h_t})^2.
\end{eqnarray}

\section{\label{sec:level3} Numerical results and conclusions}

\ \ \ In our numerical evaluation, we take a set of independent
input parameters which are known from current experiment. The input
parameters are $m_t=171.2\ {\rm GeV}$, $m_b=4.2\ {\rm GeV}$,
$\alpha=1/137.04$ and $\Gamma_t=1.377\ {\rm GeV}$ \cite{Amsler2008}.
For the c.m. energies of the ILC, we choose $\sqrt{s}=500, 1000\
{\rm GeV}$ according to the ILC Reference Design Report
\cite{Huang2009}.

\begin{figure}
\includegraphics{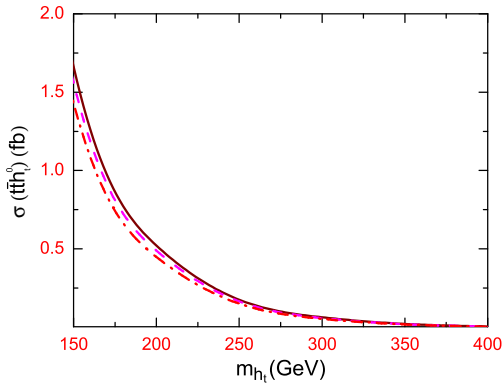}
\caption{\label{fig:eps2} The total cross section of $e^+e^-
\rightarrow \gamma\gamma \rightarrow t\bar{t}h_t$ versus $m_{h_t}$
with $\sqrt{s}=1000\ {\rm GeV}$ for $\varepsilon=0.03$ (solid),
$0.06$ (dashed) and $0.1$ (dot-dashed).}
\end{figure}

\begin{figure}
\includegraphics{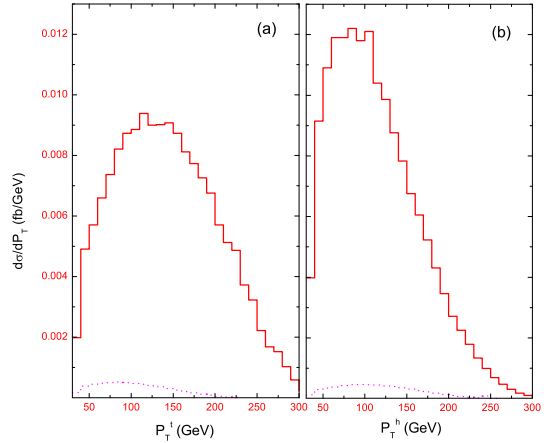}
\caption{\label{fig:eps3} The distributions of the transverse
momenta of the final states ($p^t_T$ and $p^h_T$) for the process
$\gamma\gamma \rightarrow t\bar{t}h^0_t$ at the ILC with
$\sqrt{s}=1000\ {\rm GeV}$. The solid lines and dotted lines denote
the cases of $m_{h_t} =150$ GeV and $300$ GeV, respectively.}
\end{figure}

\begin{figure}
\includegraphics{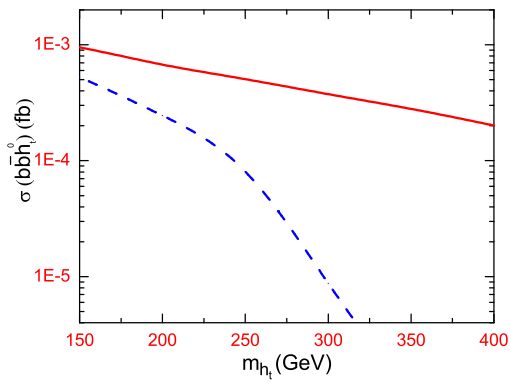}
\caption{\label{fig:eps4} The curve of $\sigma(e^+e^- \rightarrow
\gamma\gamma \rightarrow b\bar{b}h_t)$ vs. $m_{h_t}$ for
$\sqrt{s}=500\ {\rm GeV}$ (solid), $1000\ {\rm GeV}$ (dashed).}
\end{figure}

According to the idea of TC2, the masses of the first and second
generation quarks are all generated by the extended TC (ETC)
interactions. Then, the difference between $\xi_U$ and $\xi_D$ for,
respectively, the coupling coefficients techniquark to up- and
down-type quarks reflects the mass difference between the charm and
strange quarks \cite{Wu1995}. So we have $m_{t}'=(m_c/m_s)m_{b}'$,
where $m_{t}'$ and $m_{b}'$ are the top- and bottom-quark masses
generated by ETC interactions, respectively. Since $m_{b}'$  is very
small, we take approximatively $m^*_b \approx m_b =4.2\ {\rm GeV}$.
The parameter $\varepsilon$ and the mass of neutral top-Higgs $h_t$
are all model-dependent. We select them as free parameters,
$\varepsilon \sim (0.03, 0.06, 0.1$ and $150\ {\rm GeV}\leq m_{h_t}
\leq 400\ {\rm GeV}$, to estimate the total cross section of
$f\bar{f}h_t$ associated production in the high energy photon
collisions at the ILC. The final numerical results are summarized in
Figs. \ref{fig:eps2}-\ref{fig:eps4}.

The cross section $\sigma (e^+e^- \rightarrow \gamma\gamma
\rightarrow t\bar{t}h_t)$ versus the parameter $m_{h_t}$ for various
values $\varepsilon$ when $\sqrt{s} = 1000$ GeV is given in Fig.
\ref{fig:eps2}. Because for $t\bar{t}h_t$ production the c.m. energy
$\sqrt{s}=500\ {\rm GeV}$ is too low to produce it, we only consider
the case of $\sqrt{s}=1000\ {\rm GeV}$. From this figure, we can see
that: (i) the cross section decreases rapidly as $m_{h_t}$
increases. This is natural since the phase space is depressed
strongly by large $m_{h_t}$; (ii) the decrease in the cross section
is slight, with $\varepsilon$ from $0.03$, $0.06$ to $0.1$; and
(iii) the maximum of the cross section reaches the level of a few
${\rm fb}$ when $m_{h_t} \approx 150\ {\rm GeV}$.

In Fig. \ref{fig:eps3} (a) and (b), we plot the distributions of the
transverse momenta of the final states ($p^t_T$ and $p^h_T$) for the
process $\gamma\gamma \rightarrow t\bar{t}h^0_t$ at the ILC. Because
of the $CP$ conservation, the distributions of the transverse
momentum of anti-top quark, $p^{\bar{t}}_T$, should be the same as
that of ${\rm d} \sigma / {\rm d} P^t_T$ shown in Fig. 3 (a). From
these two diagrams, we can find that, (i) when $m_{h_t} =150$ GeV,
the signs of associated production $t\bar{t}h^0_t$ in the regions
around $p^t_T \sim 140$ GeV and $p^h_T \sim 100$ GeV are more
significant than in other regions, (i) when $m_{h_t} =300$ GeV, the
significant regions of associated production $t\bar{t}h^0_t$ are
around $p^t_T \sim 80$ GeV and $p^h_T \sim 90$ GeV. However, their
widths are rather large but their values are quite small.

Fig. \ref{fig:eps4} gives the results of another associated
production $b\bar{b}h_t$ in $\gamma\gamma$ collisions. We find that
the cross section of $b\bar{b}h_t$ production is much smaller than
that of $t\bar{t}h_t$ production and is only of the order of
$10^{-3} \sim 10^{-4}\ {\rm fb}$. Therefore, it is difficult to
detect indirectly $h_t$ via the process $\gamma\gamma \rightarrow
b\overline{b}h_t$ at the ILC.

We know that the ILC is the important next generation linear
collider. According to the ILC Reference Design Report
\cite{Huang2009}, the ILC is determined to run with $\sqrt{s}=500$
GeV (upgradeable to $1000$ GeV), and the total luminosity required
is $L=500\ {\rm fb}^{-1}$ for the first four-year operation and
$L=1000\ {\rm fb}^{-1}$ during the first phase of operation with
$\sqrt{s}=500$ GeV. It means that hundreds of to thousands of $h_t$
per year can be produced in high energy photon collisions at the
ILC.

The cross section of the Higgs boson in the SM associated production
with a pair of top quarks in the high energy photon collisions is at
the level of a few ${\rm fb}$ \cite{Cheung1993}. The study of the
process $\gamma\gamma \rightarrow t\overline{t}\phi$
$(\phi=h^{0},H^{0},A^{0})$ in the MSSM shows that the associated
$h^{0}$ production is dominant when $\tan \beta$ is not too large,
with the cross section of $1\ {\rm fb}$ or higher for the favorable
parameters \cite{Gunion1993}. For the effects of the little Higgs
models on $t\bar{t}h^0$ production via $\gamma\gamma$ collision at
linear colliders, the authors of Ref. \cite{Pan2007} studied the
associated $t\bar{t}h^0$ production process $e^+e^- \rightarrow
\gamma\gamma \rightarrow t\bar{t}h^0$ at the future $e^+e^-$ linear
colliders up to QCD next-to-leading order in the frameworks of the
littlest Higgs (LH) model and its extension with T-parity (LHT), and
presented the regions of $\sqrt{s}-f$ parameter space and the
production rates of process $\gamma\gamma \rightarrow t\bar{t}h^0$
in different photon polarization collision modes, and drew a
conclude that one could observe the effects contributed by the LH or
LHT model on the cross section for the process $e^+e^- \rightarrow
\gamma\gamma \rightarrow t\bar{t}h^0$ in a reasonable parameter
space, or might put more stringent constraints on the LH/LHT
parameters in the future experiments at linear colliders.

Because $h_t$ is also a scalar particle, its cross section is
basically consistent with the results of the cross section of
$t\bar{t}H$ in the SM and the cross section of $t\bar{t}h$ in the
MSSM at the level of a few ${\rm fb}$. Therefore, if the ILC
experiment could detect a scalar particle, we need to affirm which
model it is from, and it will further require more experimental data
and theoretical analysis.

In summary, we have studied the associated production processes of a
neutral top-Higgs in the TC2 model with a pair of heavy quarks in
$\gamma\gamma$ collisions at the ILC. We find that the cross section
for $t\bar{t}h_t$ in $\gamma\gamma$ collisions is at the level of a
few ${\rm fb}$ with the c.m. energy $\sqrt{s}=1000$ GeV, which
coincides with the results of the cross section of $t\bar{t}H$ in
the SM and the cross section of $t\bar{t}h$ in the MSSM. It should
be clearly visible for hundreds to thousands of $h_t$ per year
produced by the ILC, so the process of $\gamma \gamma \rightarrow
t\bar{t}h_t$ is of great interest in testing the standard model and
searching the signs of technicolor. Certainly, we need more evidence
in order to affirm the existence of $h_t$.

\section*{ACKNOWLEDGMENTS}

This project was supported in part by the National Natural Science
Foundation  of China under Grant Nos. 10975047, and 10979008; the
Natural Science Foundation of Henan Province under No. 092300410205.

\end{document}